\documentclass{aa}
\usepackage{amssymb}
\usepackage{amsmath}
\usepackage{txfonts}
\usepackage{graphicx}
\usepackage{natbib}
\bibpunct{(}{)}{;}{a}{}{,} 

\begin{document}

\title{\textsl{INTEGRAL}-\textsl{RXTE} observations of Cygnus
  X-1\thanks{Based on observations with \textsl{INTEGRAL}, an ESA
  project with instruments and science data centre funded by ESA
  member states (especially the PI countries: Denmark, France,
  Germany, Italy, Switzerland, Spain), Czech Republic and Poland, and
  with the participation of Russia and the USA.}}

\author{Katja~Pottschmidt\inst{1,2}
\and J\"orn~Wilms\inst{3}
\and Masha~Chernyakova\inst{2,4}
\and Michael~A.~Nowak\inst{5}
\and J\'er\^ome~Rodriguez\inst{6,2} 
\and Andrzej~A.~Zdziarski\inst{7}
\and Volker~Beckmann\inst{3,2}
\and Peter~Kretschmar\inst{1,2}
\and Thomas~Gleissner\inst{3}
\and Guy~G.~Pooley\inst{8}
\and Silvia~Mart\'\i{}nez-N\'u\~nez\inst{9,10}
\and \mbox{Thierry~J.-L.~Courvoisier}\inst{2,11}
\and Volker~Sch\"onfelder\inst{1}
\and R\"udiger~Staubert\inst{3}}
\institute{
   Max-Planck-Institut f\"ur extraterrestrische
   Physik, Postfach 1312, 85748 Garching, Germany
\and INTEGRAL Science
   Data Centre, Chemin d'\'Ecogia 16, 1290 Versoix, Switzerland
\and Institut f\"ur Astronomie und Astrophysik -- Abt.~Astronomie,
   Universit\"at T\"ubingen, Sand 1, 72076 T\"ubingen, Germany
\and Astro Space Center of the P.N.~Lebedev Physical Institute,
84/32 Profsoyuznaya Street, Moscow 117997, Russia
\and MIT-CXC, NE80-6077, 77 Massachusetts Ave., Cambridge, MA 02139, USA
\and DSM/DAPNIA/Service d'Astrophysique (CNRS FRE 2591), CEA Saclay,
91191 Gif-sur-Yvette, France
\and Nicolaus Copernicus Astronomical Center, Bartycka 18, 00-716 Warszawa, Poland
\and Astrophysics, Cavendish Laboratory, Madingley Road, Cambridge
CB3~0HE, United Kingdom
\and GACE, Instituto de Ciencia de los Materiales, Universidad de Valencia,
P.O. Box 22085, 46071 Valencia, Spain
\and Danish Space Research Institute, Juliane Maries Vej 30, 2100
Copenhagen \O., Denmark 
\and Observatoire de Gen\`eve, Chemin des Maillettes 51, 1290
     Sauverny, Switzerland}

\mail{K.~Pottschmidt (katja.pottschmidt@obs.unige.ch)}
\titlerunning{\textsl{INTEGRAL}-\textsl{RXTE} Observations of Cygnus~X-1}
\authorrunning{K.~Pottschmidt et al.}
\date{Received / Accepted}

\abstract{We present first results from contemporaneous observations
  of Cygnus~X-1 with \textsl{INTEGRAL} and \textsl{RXTE}, made during
  \textsl{INTEGRAL}'s performance verification phase in 2002~November
  and December. Consistent with earlier results, the 3--250\,keV data
  are well described by Comptonization spectra from a Compton corona
  with a temperature of $kT\sim50$--90\,keV and an optical depth of
  $\tau\sim 1.0$--1.3 plus reflection from a cold or mildly ionized slab
  with a covering factor of $\Omega/2\pi\sim 0.2$--0.3. A soft excess
  below 10\,keV, interpreted as emission from the accretion disk, is
  seen to decrease during the 1.5~months spanned by our observations.
  Our results indicate a remarkable consistency among the
  independently calibrated detectors, with the remaining issues being
  mainly related to the flux calibration of \textsl{INTEGRAL}.
  
  \keywords{black hole physics -- stars: individual: Cyg X-1 -- gamma
    rays: observations -- X-rays: binaries -- X-rays:
    general}}

\maketitle

\section{Introduction}\label{sec:intro}
The broad band X-ray continuum of Galactic black holes (BH) in the
hard state is characterized by a power-law with photon index
$\Gamma\sim1.7$, and an exponential cutoff at $\sim$150\,keV, which is
generally attributed to Comptonization \citep{sunyaev:79a}. A soft
excess below $\sim$1\,keV, a hardening above 10\,keV attributed to
Compton reflection, and a fluorescent 6.4\,keV Fe K$\alpha$ line are
all taken as evidence for the presence of a (possibly mildly ionized)
accretion disk. Long-term monitoring programs and the existence of
state changes from the hard state to the soft state \citep[see,
e.g.,][]{remillard:01a} show that the relative contribution of the
Comptonizing medium and the accretion disk to the emitted X-ray
luminosity are variable on long timescales
\citep{pottschmidt:02a,zdz:02a}.

The knowledge of the spectral shape above 10\,keV is crucial for the
determination of the temperature and optical depth of the Comptonizing
medium, as well as for the determination of the reflection fraction.
With the launch of the \textsl{International Gamma-Ray Astrophysics
  Laboratory} (\textsl{INTEGRAL}), a new instrument providing data
from 5\,keV to several MeV has become available. To test the
\textsl{INTEGRAL} instruments and further our knowledge of  BH
spectra, we organized simultaneous measurements of the canonical BH
\object{Cygnus~X-1} with the \textsl{Rossi X-ray Timing Explorer}
(\textsl{RXTE}) during \textsl{INTEGRAL}'s performance verification
(PV) phase. Here, we present first results of these observations. In
Sect.~\ref{sec:data} we describe the observations and the data
extraction, followed in Sect.~\ref{sec:modeling} by the results of our
modeling.  Sect.~\ref{sec:concl} summarizes our main points.

\section{Observations and data analysis}\label{sec:data}

\begin{table}
\caption{Log of the simultaneous
  \textsl{INTEGRAL}/\textsl{RXTE} observations, showing the
  \textsl{INTEGRAL} revolution (Rev.), the date of the observation, 
  as well as the exposure time of the instruments used in the data
  analysis in ksec (for \textsl{RXTE}, we give the PCA exposure
  and the dead-time corrected HEXTE live-time).}\label{tab:log} 
\begin{tabular}{ccrrrrrr}
     &          &       & \multicolumn{2}{c}{JEM} & & \\
Rev. &  Date    &  \textsl{RXTE} &X1&X2&ISGRI &   SPI \\
\hline
\hline
11 & 2002-11-16 & 12.1/8.1  &   -- &   -- & 11.7 & 86.0  \\
14 & 2002-11-26 &  6.7/4.1  &   -- & 15.0 &  0.6 &   --  \\
16 & 2002-12-02 &  4.1/2.6  &   -- & 10.1 &  5.8 &   --  \\
18 & 2002-12-06 &  7.5/5.2  & 10.6 &   -- & 12.2 &   --  \\
25 & 2002-12-29 & 29.7/34.9 &   -- &   -- &  8.0 & 86.4  \\
\end{tabular}
\end{table}

Cyg~X-1 was observed during the PV phase of \textsl{INTEGRAL} in a
variety of different observing modes. \textsl{RXTE} observing times
were chosen to coincide with \textsl{INTEGRAL} observations that
pointed at the source (satellite revolutions~11, 14,
16, and~18) as well as with observations during revolution~25, where
Cyg~X-1 was on average $\sim$$10^\circ$ off-axis. During
revolutions~14, 16, and~18 \textsl{INTEGRAL} performed staring
observations dedicated to the calibration of the Joint European X-ray
Monitor \citep[JEM-X;][]{lund:03a} and the Imager on-board
\textsl{INTEGRAL} \citep[IBIS;][]{ubertini:03a}. For both instruments,
data simultaneous to \textsl{RXTE} were extracted on 'science window
timescales' (i.e., stable \textsl{INTEGRAL} pointings of $\sim$2000\,s length).
For these revolutions, no source spectrum could be extracted from the
Spectrometer on-board \textsl{INTEGRAL} \citep[SPI;][]{vedrenne:03a},
as \textsl{INTEGRAL} was not in one of its dithering modes and the
SPI imaging solution is underdetermined.  Information from SPI is only
available for the hexagonal dithering mode in revolution~11 and for
the $5\times5$ dither pattern of revolution~25 (see
\citealt{courvoisier:03a} for a description of \textsl{INTEGRAL}'s
dithering patterns). Due to the small effective area of SPI, we use
data from the whole \textsl{INTEGRAL} revolution contemporaneous to
the \textsl{RXTE} observations. For the IBIS analysis of these two
revolutions, only the closest contemporaneous science windows were
considered. These were either on-axis (rev.~11) or had an offset of
$<5^\circ$ (rev.~25). The JEM-X data extraction failed for these two
revolutions. Table~\ref{tab:log} provides a log of all the
observations analyzed herein.  See \citet{bazzano:03a} and
\citet{laurent:03a} for discussions of (non-simultaneous with \textsl{RXTE}) 
IBIS data of Cyg X-1.

The \textsl{INTEGRAL} data were extracted using the \textsl{INTEGRAL}
off-line scientific analysis (OSA) software, version 1.1, including
those fixes to the released software which were available up to
2003~June~15. For all instruments, the newest available response
matrices were used. Data from JEM-X were considered from 10--40\,keV
only, due to on-board event selection effects in the early phase of
the mission (before rev.~45). For IBIS we used data from 40--250\,keV,
the energy band in which the energy response of IBIS was considered
calibrated at the time of writing, and only considered the
\textsl{INTEGRAL} Soft Gamma-Ray Imager (ISGRI) part of the detector.
We applied the 10\% systematic uncertainty recommended by the IBIS
team (Ubertini, priv.\ comm.) in quadrature to the data. For SPI, data
from 50--200\,keV were extracted with SPIROS, version 4.3.4
\citep{skinner:03a,strong:03a}. See Sect.~\ref{sec:modeling} for a
discussion of this rather low upper energy threshold of SPI.  For all
instruments, background subtraction has been taken into account by the
OSA. Current experience shows, e.g., that with ISGRI a few ks are
sufficient to extract a significant spectrum up to $\sim$100\,keV for
a 60\,mCrab source \citep{paizis:03a}, thus background influences
should not be an issue for Cyg X-1 up to a few 100\,keV.

\begin{figure}
\resizebox{\hsize}{!}{\includegraphics{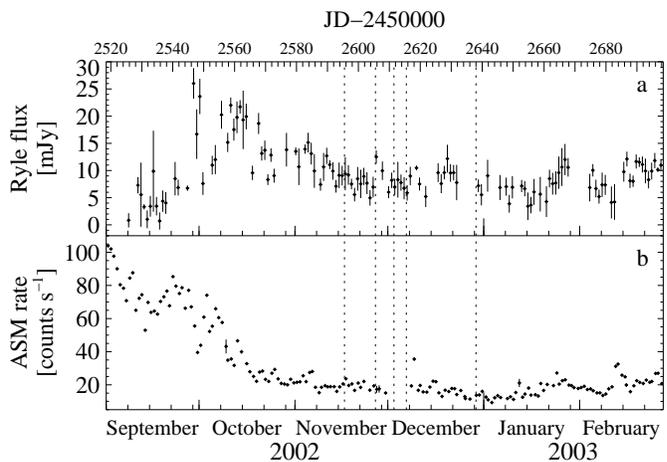}}
\caption{\textbf{a} 15\,GHz radio lightcurve from the Ryle telescope
and \textbf{b} 2--12\,keV \textsl{RXTE}-All Sky Monitor flux of
Cyg~X-1 for the time of the pointed \textsl{INTEGRAL} and
\textsl{RXTE} observations (dashed lines). Note the end of the 2002
soft state of Cyg~X-1 in 2002 mid-October.}\label{fig:longterm}
\end{figure}

We compare the \textsl{INTEGRAL} data with measurements from both
\textsl{RXTE} pointed instruments, the Proportional Counter Array
(PCA; \citealt{jahoda:96b}) and the High Energy X-ray Timing
Experiment \citep[HEXTE;][]{rothschild:98a}. The \textsl{RXTE} data
were extracted with the HEASOFT software, version~5.2, using our
standard procedures \citep[e.g.,][]{wilms:98c}. The high effective
area of the PCA ($\sim$4000\,$\text{cm}^2$) necessitates the use of
energy dependent systematic uncertainties of 1\% for PCA channels
0--15 ($\lesssim$7.5\,keV), 0.5\% for channels 16--39
(7.5\,keV--18\,keV), and 2\% up to 25\,keV \citep{kreykenbohm:03a}.
These uncertainties are added in quadrature to the data. For the
HEXTE, the 20\,keV--200\,keV energy range was considered.  Finally,
the Ryle telescope in Cambridge, England, performed simultaneous radio
observations at 15\,GHz in addition to its daily monitoring of the
source (Fig.~\ref{fig:longterm}).

Spectral fitting was performed with XSPEC~11.2.0bc \citep{arnaud:96a},
which includes important fixes to the reflection model used in the
spectral analysis (Sect.~\ref{sec:modeling}). The uncertainty of the
instrument flux calibration was taken into account by introducing a
multiplicative constant into the spectral models, and normalizing all
fluxes to the PCA. This constant was 0.80--0.84 for the HEXTE
(consistent with earlier results), 0.56 for JEM-X1, 0.43 and 0.46 for
JEM-X2, 1.03 and 1.05 for SPI, and it varied between 0.89 and 1.38 for
ISGRI (see Tables~\ref{tab:comptt} and~\ref{tab:eqpair} below). The
latter large variation is mainly due to known issues with the current
ISGRI deadtime correction.

\section{Modeling the 3--250\,keV spectrum}\label{sec:modeling}

\begin{figure*}
\centering
\includegraphics[width=17cm]{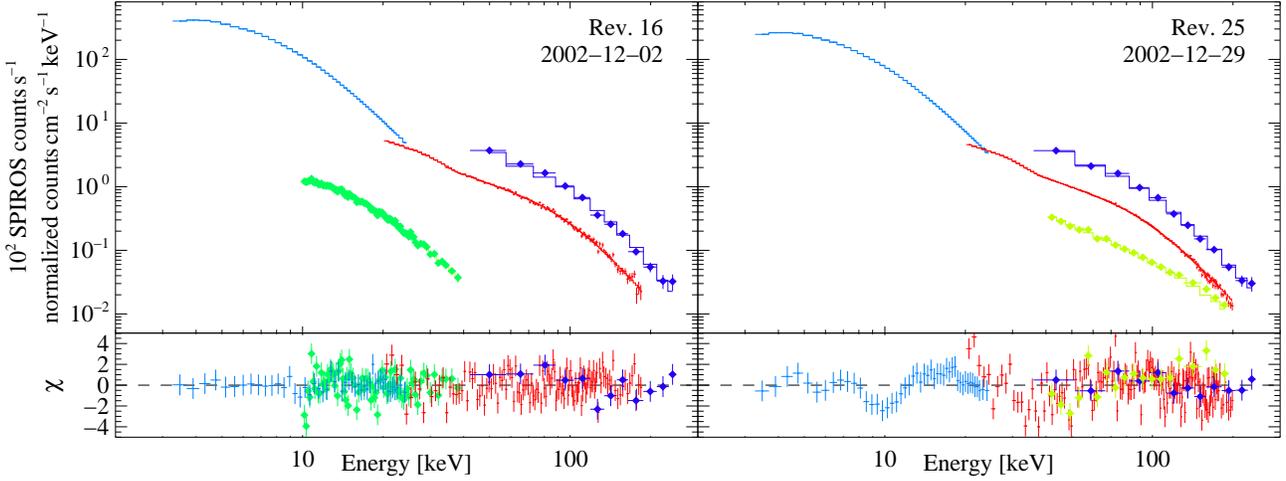}
\caption{Spectra, best fit \texttt{compTT} model,  and residuals for
  the observations of 2002 December~02 and 2002 December~29 (PCA,
  3--25\,keV, light blue; JEM-X, 10--40\,keV, green; HEXTE,
  20--200\,keV, red; IBIS, 40--250\,keV, dark blue; SPI, 50--200\,keV,
  olive green). For SPI, the spectrum as provided by the SPI spectral
  extractor, SPIROS, is shown, for the other instruments, we display
  the count rate normalized by the detector effective area.  The PCA
  residuals for revolution~25 are consistent with known systematic
  effects.}\label{fig:ldata}
\end{figure*}

\begin{figure*}
\centering
\includegraphics[width=17cm]{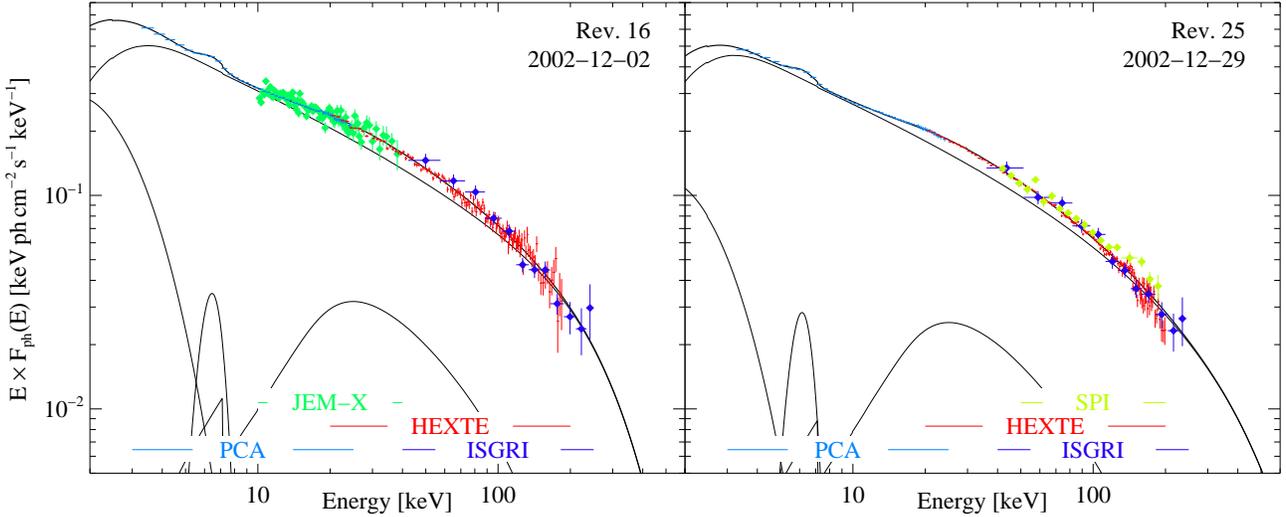}
\caption{Unfolded X-ray spectra for the observations of
  Fig.~\ref{fig:ldata}, and the components of the \texttt{compTT}
  model.  For all instruments, the measured fluxes have been
  renormalized to that measured by the PCA, the colors are those of
  Fig.~\ref{fig:ldata}.}\label{fig:ufspec}
\end{figure*}

\begin{table*}
\caption{Best fit parameters for the \texttt{compTT} model (full model
in XSPEC notation:
\texttt{const$\times$phabs[diskbb+gauss+compTT+reflect(comptt)]}). The
parameters shown are the inner accretion disk temperature
$kT_\text{in}$ and its normalization,
$A_\text{disk}=((R_\text{in}/\text{km})/(D/10\,\text{kpc}))^2\cos i$,
the electron temperature of the Comptonizing plasma, $kT_\text{e}$ and
its optical depth $\tau$, the covering factor of the cold reflecting
medium, $\Omega/2\pi$, and the energy and width of the Gaussian Fe
K$\alpha$ line, $E_\text{K$\alpha$}$ and $\sigma_\text{K$\alpha$}$,
and the flux normalization constants of the individual instruments
with respect to the PCA are given by $c_\text{HEXTE}$,
$c_\text{ISGRI}$, $c_\text{JEM-X1}$, $c_\text{JEM-X2}$, and
$c_\text{SPI}$. For revolutions 11 and 25 the results are shown with
and without taking SPI into account in order to 
illustrate its systematic influences.}\label{tab:comptt}
\begin{tabular}{lrrrrrrr}
 & Rev.~11 & Rev.~14 & Rev.~16 & Rev.~18 & Rev.~25 & Rev.~11 & Rev.~25\\
 &         &         &         &         &         &     SPI &     SPI\\
\hline
\multicolumn{1}{l}{$kT_\text{in}$ [eV] } & $735^{+77}_{-113}$ & $745^{+102}_{-162}$ & $710^{+72}_{-122}$ & $772^{+90}_{-157}$ & $660^{+65}_{-143}$ & $670^{+87}_{-90}$ & $651^{+75}_{-119}$ \\[0.3mm]
\multicolumn{1}{l}{$A_\text{disk}$} & $540^{+790}_{-200}$ & $460^{+1140}_{-190}$ & $440^{+800}_{-150}$ & $430^{+820}_{-150}$ & $260^{+300}_{-80}$ & $650^{+120}_{-50}$ & $270^{+450}_{-90}$ \\[0.3mm]
\multicolumn{1}{l}{$kT_\text{e}$ [keV] } & $62^{+7}_{-8}$ & $54^{+7}_{-4}$ & $62^{+19}_{-9}$ & $51^{+3}_{-3}$ & $62^{+4}_{-4}$ & $252^{+26}_{-32}$ & $94^{+39}_{-20}$ \\[0.3mm]
\multicolumn{1}{l}{$\tau$} & $1.01^{+0.15}_{-0.11}$ & $1.21^{+0.09}_{-0.13}$ & $1.06^{+0.17}_{-0.26}$ & $1.24^{+0.07}_{-0.07}$ & $1.07^{+0.08}_{-0.06}$ & $0.16^{+0.05}_{-0.04}$ & $0.69^{+0.20}_{-0.17}$ \\[0.3mm]
\multicolumn{1}{l}{$\Omega/2\pi$} & $0.17^{+0.01}_{-0.01}$ & $0.17^{+0.01}_{-0.01}$ & $0.19^{+0.01}_{-0.02}$ & $0.19^{+0.01}_{-0.01}$ & $0.17^{+0.01}_{-0.01}$ & $0.24^{+0.01}_{-0.01}$ & $0.17^{+0.00}_{-0.00}$ \\[0.3mm]
\multicolumn{1}{l}{$E_\text{K$\alpha$}$ [keV] } & $6.42^{+0.22}_{-0.39}$ & $6.28^{+0.31}_{-0.65}$ & $6.38^{+0.25}_{-0.49}$ & $6.24^{+0.25}_{-0.54}$ & $6.13^{+0.27}_{-0.35}$ & $6.44^{+0.13}_{-0.13}$ & $6.10^{+0.30}_{-0.28}$ \\[0.3mm]
\multicolumn{1}{l}{$\sigma_\text{K$\alpha$}$ [keV] } & $0.74^{+0.32}_{-0.31}$ & $0.86^{+0.41}_{-0.39}$ & $0.66^{+0.39}_{-0.35}$ & $0.78^{+0.39}_{-0.35}$ & $0.59^{+0.32}_{-0.31}$ & $0.61^{+0.15}_{-0.14}$ & $0.62^{+0.31}_{-0.32}$ \\[0.3mm]
\multicolumn{1}{l}{$c_\text{HEXTE}$} & $0.83^{+0.01}_{-0.00}$ & $0.81^{+0.01}_{-0.01}$ & $0.80^{+0.01}_{-0.01}$ & $0.84^{+0.01}_{-0.01}$ & $0.83^{+0.01}_{-0.00}$ & $0.82^{+0.00}_{-0.00}$ & $0.83^{+0.00}_{-0.00}$ \\[0.3mm]
\multicolumn{1}{l}{$c_\text{ISGRI}$} & $1.08^{+0.05}_{-0.05}$ & $0.89^{+0.06}_{-0.06}$ & $1.36^{+0.06}_{-0.07}$ & $1.38^{+0.05}_{-0.06}$ & $1.23^{+0.05}_{-0.05}$ & $1.01^{+0.04}_{-0.04}$ & $1.21^{+0.05}_{-0.05}$ \\[0.3mm]
\multicolumn{1}{l}{$c_\text{JEM-X1}$} &  --  &  --  &  --  & $0.560^{+0.004}_{-0.004}$ &  --  &  --  &  --  \\[0.3mm]
\multicolumn{1}{l}{$c_\text{JEM-X2}$} &  --  & $0.456^{+0.003}_{-0.003}$ & $0.428^{+0.004}_{-0.004}$ &  --  &  --  &  --  &  --  \\[0.3mm]
\multicolumn{1}{l}{$c_\text{SPI}$} &  --  &  --  &  --  &  --  &  --  & $1.05^{+0.00}_{-0.01}$ & $1.03^{+0.01}_{-0.01}$ \\[0.3mm]
\multicolumn{1}{l}{$\chi^2/\text{dof}$} & $228$/$240$ & $321$/$317$ & $387$/$318$ & $352$/$339$ & $505$/$240$ & $413$/$279$ & $589$/$253$ \\[0.3mm]
\multicolumn{1}{l}{$\chi^2_\text{red}$} &  0.95 &  1.01 &  1.22 &  1.04 &  2.11 &  1.48 &  2.33 \\[0.3mm]
\end{tabular}

\end{table*}

\begin{table*}
\caption{Best fit parameters for the \texttt{eqpair} fits (full model
in XSPEC notation:
\texttt{const$\times$phabs[diskbb+gauss+eqpair]}). Shown are the disk
parameters $kT_\text{in}$ and $A_\text{disk}$, inner accretion disk
temperature, $kT_\text{in}$, the ratio of the compactness of the
Comptonizing plasma to the accretion disk compactness,
$l_\text{h}/l_\text{s}$, from which the Comptonization temperature
$kT_\text{e}$ can be derived (as $kT_\text{e}$ is not a fit parameter,
we do not give its uncertainty), the optical depth of the Comptonizing
plasma, $\tau$, the covering factor of the reflecting medium,
$\Omega/2\pi$, and its ionization parameter, $\xi$, the energy
$E_\text{K$\alpha$}$ and width $\sigma$ of the Fe K$\alpha$ line
(where $E_\text{K$\alpha$}$ was limited to $\ge$6\,keV in the fits), and
the flux normalization constants of the instruments with respect to
the PCA. As in Tab.~\ref{tab:comptt}, for revolutions 11 and 25 the
results are shown with and without taking SPI into
account.}\label{tab:eqpair} 
\begin{tabular}{lrrrrrrr}
 & Rev.~11 & Rev.~14 & Rev.~16 & Rev.~18 & Rev.~25 & Rev.~11 & Rev.~25\\
 &         &         &         &         &         &     SPI &     SPI\\
\hline
\multicolumn{1}{l}{$kT_\text{in}$ [eV] } & $256^{+1}_{-5}$ & $250^{+5}_{-8}$ & $249^{+6}_{-8}$ & $251^{+5}_{-6}$ & $274^{+9}_{-44}$ & $258^{+3}_{-7}$ & $254^{+25}_{-29}$ \\[0.3mm]
\multicolumn{1}{l}{$A_\text{disk}/10^5$} & $36^{+8}_{-7}$ & $32^{+9}_{-9}$ & $23^{+9}_{-9}$ & $34^{+5}_{-9}$ & $0^{+1}_{-0}$ & $38^{+6}_{-7}$ & $0^{+2}_{-0}$ \\[0.3mm]
\multicolumn{1}{l}{$l_\text{h}/l_\text{s}$} & $9.0^{+0.1}_{-0.1}$ & $10.0^{+0.1}_{-0.1}$ & $10.0^{+0.2}_{-0.2}$ & $9.7^{+0.1}_{-0.1}$ & $10.2^{+0.1}_{-0.1}$ & $8.8^{+0.1}_{-0.1}$ & $10.5^{+0.1}_{-0.2}$ \\[0.3mm]
\multicolumn{1}{l}{$kT_\text{e}$ [keV] } & $85$ & $85$ & $85$ & $73$ & $89$ & $146$ & $89$ \\[0.3mm]
\multicolumn{1}{l}{$\tau$} & $1.29^{+0.03}_{-0.03}$ & $1.33^{+0.02}_{-0.04}$ & $1.33^{+0.05}_{-0.06}$ & $1.52^{+0.01}_{-0.01}$ & $1.31^{+0.02}_{-0.02}$ & $0.75^{+0.01}_{-0.03}$ & $1.30^{+0.01}_{-0.02}$ \\[0.3mm]
\multicolumn{1}{l}{$\Omega/2\pi$} & $0.26^{+0.01}_{-0.01}$ & $0.25^{+0.01}_{-0.01}$ & $0.27^{+0.02}_{-0.01}$ & $0.26^{+0.01}_{-0.01}$ & $0.28^{+0.01}_{-0.01}$ & $0.28^{+0.01}_{-0.01}$ & $0.28^{+0.01}_{-0.01}$ \\[0.3mm]
\multicolumn{1}{l}{$\xi$} & $89^{+11}_{-49}$ & $0^{+4}_{-0}$ & $43^{+53}_{-38}$ & $0^{+8}_{-0}$ & $32^{+23}_{-16}$ & $0^{+22}_{-0}$ & $32^{+21}_{-17}$ \\[0.3mm]
\multicolumn{1}{l}{$E_\text{K$\alpha$}$ [keV] } & $6.00^{+0.11}_{-0.00}$ & $6.00^{+0.07}_{-0.00}$ & $6.00^{+0.53}_{-0.00}$ & $6.00^{+2.00}_{-0.00}$ & $6.29^{+0.24}_{-0.29}$ & $6.00^{+0.03}_{-0.00}$ & $6.30^{+0.24}_{-0.30}$ \\[0.3mm]
\multicolumn{1}{l}{$\sigma_\text{K$\alpha$}$ [keV] } & $0.86^{+0.19}_{-0.29}$ & $0.98^{+0.14}_{-0.14}$ & $0.78^{+0.30}_{-0.73}$ & $0.92^{+1.15}_{-0.13}$ & $0.29^{+0.40}_{-0.29}$ & $1.00^{+0.06}_{-0.16}$ & $0.29^{+0.40}_{-0.29}$ \\[0.3mm]
\multicolumn{1}{l}{$c_\text{HEXTE}$} & $0.84^{+0.00}_{-0.18}$ & $0.82^{+0.01}_{-0.14}$ & $0.82^{+0.01}_{-0.01}$ & $0.85^{+0.10}_{-0.00}$ & $0.82^{+0.00}_{-0.00}$ & $0.83^{+0.00}_{-0.00}$ & $0.82^{+0.15}_{-0.00}$ \\[0.3mm]
\multicolumn{1}{l}{$c_\text{ISGRI}$} & $1.10^{+0.05}_{-0.04}$ & $0.90^{+0.06}_{-0.06}$ & $1.38^{+0.06}_{-0.06}$ & $1.39^{+0.05}_{-0.05}$ & $1.23^{+0.05}_{-0.05}$ & $1.03^{+0.04}_{-0.04}$ & $1.23^{+0.05}_{-0.05}$ \\[0.3mm]
\multicolumn{1}{l}{$c_\text{JEM-X1}$} &  --  &  --  &  --  & $0.561^{+0.004}_{-0.004}$ &  --  &  --  &  --  \\[0.3mm]
\multicolumn{1}{l}{$c_\text{JEM-X2}$} &  --  & $0.456^{+0.003}_{-0.003}$ & $0.429^{+0.004}_{-0.004}$ &  --  &  --  &  --  &  --  \\[0.3mm]
\multicolumn{1}{l}{$c_\text{SPI}$} &  --  &  --  &  --  &  --  &  --  & $1.06^{+0.01}_{-0.01}$ & $1.04^{+0.01}_{-0.01}$ \\[0.3mm]
\multicolumn{1}{l}{$\chi^2/\text{dof}$} & $246$/$239$ & $335$/$316$ & $380$/$317$ & $370$/$338$ & $406$/$239$ & $440$/$278$ & $509$/$252$ \\[0.3mm]
\multicolumn{1}{l}{$\chi^2_\text{red}$} &  1.03 &  1.06 &  1.20 &  1.10 &  1.70 &  1.58 &  2.02 \\[0.3mm]
\end{tabular}

\end{table*}

We modeled the spectra using the canonical model for the hard state, a
Comptonization continuum which is partly reflected off a slab-like
accretion disk. In these models, it is assumed that the Comptonizing
medium sandwiches the accretion disk, although its covering factor can
be less than unity.  This geometry is not the only possible one for
Cyg~X-1.  Sphere+disk models, e.g., where the Comptonizing medium is
an inner spherical accretion flow irradiated by an outer accretion
disk (possibly slightly penetrating the sphere) have also been shown
to describe the data \citep{dove:97b,dove:97c,zdz:99a}.  Furthermore,
emission from the base of a jet might also contribute to the observed
X-rays \citep{markoff:02a}. We will consider such more complex models
in a future paper.

Due to the limitations intrinsic to the available Comptonization
models, it is generally useful to apply several different
Comptonization models to the data. Such an approach allows one to test
whether any trends seen in the evolution of spectral fit parameters
are stable against these limitations. See, e.g., \citet{nowak:02a} for
a discussion of these issues in the context of \textsl{RXTE}
observations of \object{GX~339$-$4}.  We therefore describe our joint
\textsl{INTEGRAL}-\textsl{RXTE} spectra using two different kinds of
spectral models. The first one is based on the theory of
\citet[][XSPEC model \texttt{compTT}]{hua:95a} as a Comptonization
continuum, which assumes a Wien spectrum for the seed photons. This
continuum is then partly reflected off a cold accretion disk
\citep{magdziarz:95a}. To account for the effects of a patchy
Comptonizing plasma, an additional soft disk black body
\citep{mitsuda:84a} is added. The temperature at the inner edge of the
disk, $kT_\text{in}$, is set equal to the seed photon temperature. As
the reflection model does not include fluorescent line emission, the
Fe K$\alpha$ line at 6.4\,keV is described by a Gaussian. We also use
a fixed absorbing column of $N_\text{H}=6\times
10^{21}\,\text{cm}^{-2}$ \citep{balu:95a}, as the PCA 3\,keV lower
threshold does not allow such low columns to be constrained.

We also described our data using the thermal Comptonization model of
\citet[][XSPEC modelname \texttt{eqpair}]{coppi:99a}\footnote{The
  \texttt{eqpair}-model is available as a local XSPEC model at
  \texttt{http://www.astro.yale.edu/coppi/}.}, which includes
reflection off a possibly ionized accretion disk
\citep{magdziarz:95a,done:92a}.  In this case, the spectral shape of
the seed photons is a disk black body, part of which is not
Comptonized to take into account the possibility of a patchy Compton
corona. This has again been realized in form of an additional disk
black body component of the same temperature as the seed photons. The
Fe K$\alpha$ line is again described by a Gaussian and absorption is
treated identical as in the \texttt{compTT} fits.

Both models resulted in equally good descriptions of the data. We show
examples of our best fits in Figs.~\ref{fig:ldata}
and~\ref{fig:ufspec}, and list the best fit parameters for all
observations in Tables~\ref{tab:comptt} and~\ref{tab:eqpair}.  For
revolution 25, $\chi^2_\text{red}=2.1$, for the other observations
$\chi_\text{red}^2<1.5$.  Due to the long PCA exposure of rev.~25,
$\chi^2$ is dominated by the PCA (the PCA residuals shown for
revolution 25 in Fig.~\ref{fig:ldata} are consistent with calibration
uncertainties).  We therefore conclude that all of our Comptonization
fits resulted in a satisfactory description of the 3--200\,keV data
from Cyg~X-1.

For the \texttt{compTT} model, the Comptonization parameters are in
principle consistent with earlier observations
\citep{dove:97c,gierlinski:97a}. For all five observations, the
Comptonization temperature lies between 50 and 60\,keV and the Thomson
optical depth of the corona between 1 and 1.2 (cf.\ the two last
columns of Table~\ref{tab:comptt} for systematic influences from the
SPI spectrum). These parameters vary only slightly with time, as is
expected for the hard state of Cyg~X-1. The covering factor of the
reflecting slab is $\Omega/2\pi \sim 0.18$. This rather small value
could imply that the Compton corona only covers part of the accretion
disk, or that the accretion geometry is more complicated than the slab
geometry assumed here \citep[sphere+disk models, e.g., imply
$\Omega/2\pi\lesssim 0.3$;][]{dove:97b}. The \texttt{eqpair} models
agree qualitatively with these results and also require moderate
$\tau$ and $\Omega/2\pi < 0.5$. They also allow for a very modest
amount of ionization, with ionization parameters $\xi\lesssim 100$.
For both, \texttt{compTT} and \texttt{eqpair}, iron lines were
consistent with equivalent widths of $\sim$50 to 100\,eV, and are
possibly broad.

Contrary to this constant behavior above 10\,keV, the spectrum is much
more variable on the long term at lower energies
\citep[e.g.,][Fig.~12]{zdz:02a}. Here we see a systematic evolution of
the soft X-ray spectrum.  For the \texttt{compTT} model fits, the flux
of the non-Comptonized disk component decreases by a factor of about
two between the first and the last of our observations, whereas for
the \texttt{eqpair} fits, the flux of the additional disk component
nearly completely vanishes. It is tempting to interpret this decrease
in the flux of the soft model component as a true physical indication
of the final stages of the 2002 transition out of the soft state,
given the proximity of our first observation to the end of this state
(Fig.~\ref{fig:longterm}).

The temperature of the non-Comptonized disk component is rather high
for the \texttt{compTT} fits ($kT_\text{in}\sim 700$\,eV, implying an
inner disk radius comparable to that of the last marginally stable
orbit of a disk around a $10\,\text{M}_\odot$ Schwarzschild black
hole).  There have been reports of a soft excess in the hard state of
Cyg X-1 in addition to the black body provided by the disk
\citep[e.g.,][]{disalvo:01a} and the hard Comptonized component. In
principle, such an excess might explain our \texttt{compTT}
parameters, however, our data do not allow to evaluate more complex
models for the soft energy regime. Also, the disk temperature found in
the \texttt{eqpair} fits is significantly lower.  Here it is close to
the $kT_\text{in}=200$\,eV found, e.g., with ROSAT \citep{balu:95a},
with an implied disk radius near 50\,$GM/c^2$ for a
10\,$\text{M}_\odot$ BH. The Fe K$\alpha$ line energy in the
\texttt{eqpair} fits is systematically too low, however, possibly
indicating the soft spectral complexity noted above.

This discrepancy of the temperatures and implied disk radii might be
caused by slight differences in the treatment of the transition of the
seed photon spectrum to the Comptonization spectrum, especially given
that \texttt{compTT} describes the seed photons with a simple Wien
spectrum. Note also that constraining the temperature of the disk
black body component to the seed photon temperature -- a rather
established simplification -- removes some degeneracy from the low
energy parameters. Finally, we also cannot exclude the possibility
that part of the fitted black body is due to the calibrations of the
PCA below $\sim$10\,keV, especially given that the temperatures of the
soft components for both cases, \texttt{compTT} and \texttt{eqpair}
fits, imply the peak emission is falling below the $\sim$3\,keV lower
cutoff of the PCA. We conclude that our simplifications of the low
energy spectral model are justified, especially since their effect on
the parameters derived for the spectrum above 10\,keV can be expected
to be negligible. Note that the spectral shapes measured above 10\,keV
band with the PCA, HEXTE, JEM-X, and ISGRI are indeed remarkably
consistent with each other, suggesting that their respective
calibrations are in agreement.

Measurements with the Compton Gamma-Ray Observatory (CGRO) and \textsl{RXTE}
show the presence of a hard tail from non-thermal Comptonization at
$\gtrsim$300\,keV in the hard state of Cyg~X-1 and above several keV
in the soft state \citep{gierlinski:99a,mcconnell:00a,mcconnell:01a}.
In principle, ISGRI and SPI detect Cyg~X-1 to at least 500\,keV and
should be able to detect significantly this hard tail.  Currently,
however, the intercalibration of the instruments above 250\,keV is not
yet good enough to allow an extrapolation to this energy band. While
ISGRI and HEXTE are in good agreement, showing a spectral turnover and
having a peak of $\nu f_\nu$ at $\sim$100\,keV, the SPI spectrum is
consistent with a continuation of a pure power-law.

\section{Conclusions}\label{sec:concl}
We have presented first results from fitting simultaneous
\textsl{INTEGRAL}-\textsl{RXTE} data of Cyg~X-1 with two different
Comptonization models. Our results for the spectral shape are in
basic agreement with earlier results, with HEXTE, ISGRI, JEM-X,
and to a lesser extent SPI, agreeing with each other. We take this
result to mean that the energy redistribution of the \textsl{INTEGRAL}
instruments is well understood up to $\sim$250\,keV.  The flux
normalizations of the instruments, however, still show large
deviations among instruments. This is likely caused by insufficient
modeling of the instrumental deadtimes, the deconvolution of the coded
mask images, and the calibration of the off-axis effective area.

The high total signal above 10\,keV in principle allows us to
constrain the continuum shape to a higher degree than has been
possible with \textsl{RXTE} or \textsl{BeppoSAX} alone. The
presence of an additional, comparatively weak Comptonized component
cannot be constrained. In two of the reported cases the additional
Comptonized component peaks at $\sim$1\,keV, and thus only affects the
spectrum at $\lesssim$10\,keV \citep[][\textsl{BeppoSax}
data]{disalvo:01a,frontera:01a}. \textsl{Ginga} and \textsl{OSSE} data
were analyzed using different Comptonization models from those applied
here and direct comparisons cannot be made
\citep[][kT$_1\sim110$\,keV, $\tau_1\sim1.7$, kT$_2\sim45$\,keV,
$\tau_2\sim6$]{gierlinski:97a}.  Interestingly, the use of a directly
comparable approach \citep[][\texttt{eqpair},
\textsl{RXTE}/\textsl{OSSE} data]{maccarone:03a} also results in a
good spectral description using a single thermal Comptonization
component and parameters fully consistent with our non-SPI fits. This
might indicate a good stability of the hard state spectral shape above
10\,keV on a time scale of years.

Thus it is quite clear that apart from model details our overall
description of the hard state spectrum above 10\, keV -- i.e., with
thermal Comptonization and a covering factor of the reflector of
$\Omega/2\pi < 0.3$ -- is principally the same as found previously
\citep{dove:97c,gierlinski:97a,disalvo:01a,frontera:01a,maccarone:03a}.
Comptonization plus reflection models, featuring a partial covering of
the reflector or a geometrically more complicated accretion disk
configuration, are thus still feasible explanations of the X-ray
spectrum. Direct comparisons between these models and other
alternative explanations, e.g., involving jet physics, still need to
be performed. The decrease of the observed black body flux in the PCA
data could be caused by the accretion disk fading away after the 2002
soft state.

We stress that this paper only contains the preliminary fits employing
the earliest available response models for \textsl{INTEGRAL}. These
models, however, already show a remarkable consistency among
independent detectors over a range from 10--200\,keV, with the
remaining issues being mainly related to the flux calibration of
\textsl{INTEGRAL}. Despite these limitations we have shown that
Comptonization models already provide a good description of the joint
3--250\,keV \textsl{INTEGRAL}-\textsl{RXTE} data. As the calibration
of \textsl{INTEGRAL} improves, tests using more advanced models than
those employed here will become possible.  Since the \textsl{INTEGRAL}
data will cover the important range above 250\,keV, where
Klein-Nishina effects become important for the shape of the
Comptonization continuum and where the non-thermal tails start to
dominate, it is expected in the near future that these data will allow
us to constrain the important physical parameters responsible for the
generation of the X-ray spectrum of the hard state of Galactic black
holes.

\begin{acknowledgements}
  This work has been financed by Deutsches Zentrum f\"ur Luft-
  und Raumfahrt grants 50~OG~95030 and 50~OG~9601, Deutsche
  Forschungsgemeinschaft grant Sta~173/25-3, National Science
  Foundation grant INT-0233441, National Aeronautics and Space
  Administration grant NAS8-01129, Deutscher Akademischer
  Austauschdienst grant D/0247203, KBN grants 5P03D00821,
  2P03C00619p1,2, and PBZ-054/P03/2001, as well as the French Space
  Agency (CNES) and the Foundation for Polish Science. We thank all
  people involved in building and calibrating \textsl{INTEGRAL} for
  their efforts, and E.~Smith and J.~Swank for the very smooth
  scheduling of the \textsl{RXTE} observations. We would also like to
  thank Paolo Coppi, the referee, for helpful suggestions.
\end{acknowledgements}

\end{document}